\documentclass[12pt]{article}
\usepackage{epsfig,amsfonts,amssymb}
\usepackage{amsmath,graphics}
\usepackage{slashed}
\usepackage{scrextend}

 \usepackage{hyperref}
\usepackage{cleveref}
\crefformat{footnote}{#2\footnotemark[#1]#3}
\def\ZZZ{{\hbox{ Z\kern-1.6mm Z}}}
\def\RRR{{\hbox{ R\kern-2.4mm R}}}
\def\CCC{{\hbox{ C\kern-2.0mm C}}}
\def\zzz{{\hbox{z\kern-1mm z}}}

\def\one{{\hbox{ 1\kern-.8mm l}}}
\def\zero{{\hbox{ 0\kern-1.5mm 0}}}

\begin{document}

\baselineskip 24pt

\begin{center}
{\Large \bf Classical Monopole Solutions to $\mathcal{N}=3$ Chern-Simons-Yang-Mills Theories}

\end{center}

\vskip .6cm
\medskip

\vspace*{4.0ex}

\baselineskip=18pt

\centerline{\large \rm   Bobby Ezhuthachan$^{a\,1}$, Chethan N. Gowdigere$^{b\,c\,2}$, Moumita Patra$^{b\,c\,3}$}

\vspace*{4.0ex}

\centerline{\large \it $^a$Ramakrishna Mission Vivekananda Educational and Research Institute,}

\centerline {\large \it Belur Math, Howrah 711202, West Bengal, INDIA}

\vspace*{1.0ex}

\centerline{\large \it $^b$National Institute of Science Education and Research Bhubaneshwar,}

\centerline{\large \it  P.O. Jatni, Khurdha, 752050, Odisha, INDIA}

\vspace*{1.0ex}

\centerline{\large \it $^c$Homi Bhabha National Institute,}

\centerline{\large \it Training School Complex, Anushakti Nagar, Mumbai 400094, INDIA}

\vspace*{4.0ex}
\centerline{E-mail: $^1$bobby.ezhuthachan@rkmvu.ac.in, $^2$chethan.gowdigere@niser.ac.in, $^3$mpatra91@niser.ac.in }

\vspace*{5.0ex}

\centerline{\bf Abstract} \bigskip

We construct the component action and supersymmetry transformations  of the $\mathcal{N} = 3$ Chern-Simons-Yang-Mills theory that flows in the infra red to the $\mathcal{N} = 3$ super conformal quiver Chern-Simons gauge theory obtained by Jafferis and Tomasiello in \cite{Jafferis:2008qz}. We then obtain classical $\frac{1}{3}$-BPS (and anti-BPS) solutions all along the RG flow. The solutions display a rich structure, for example, there is a non-trivial moduli space. Various properties of supersymmetric monopole operators in the infra red super conformal field theory, such as global charges and scaling dimensions can be derived from the monopole solutions we obtain in this paper; along the lines of the work of Benna, Klebanov and Klose in \cite{Benna:2009xd}.

\vfill \eject

\baselineskip=18pt

\tableofcontents

\section{\label{1}Introduction}

In this paper, we study three dimensional $\mathcal{N}=3$ Chern-Simons-matter theories that were constructed by Jafferis and Tomasiello in \cite{Jafferis:2008qz}. These theories, referred to as Jafferis-Tomasiello theories,  arose as world volume theories of $M2$ branes placed at  singularities in the transverse eight dimensional hyper-Kahler manifold and are dual by the AdS-CFT conjecture \cite{Maldacena:1997re, Witten:1998qj, Gubser:1998bc} to M-theory on $AdS_4$ times the seven dimensional tri-Sasakian manifold whose cone is the eight dimensional hyper-Kahler manifold. Jafferis-Tomasiello theories are quiver gauge theories: there are $n$ nodes each with a gauge group factor $U(N)$ and there are $n$ edges arranged in a circle. At each node there is a $\mathcal{N}=3$ vector multiplet $(\mathcal{V}_{(j)}, \Phi_{{(j)}})$; indices such as $(j)$, written within parentheses indicate this node index  which runs from $1$ to $n$ (when they are repeated in an expression, they are not summed over.) $\mathcal{V}_{(j)}$ is a $\mathcal{N}=2$ vector superfield and $\Phi_{{(j)}}$ is a $\mathcal{N}=2$ chiral superfield, both in the adjoint representation of the  $U(N)$ gauge group factor pertinent to the $(j)$'th node.  The gauge group is thus $\Pi_{(j)}\,U(N)_{(j)}$. Also there is a Chern-Simons level associated to each node $k_{(j)}$ with the  condition that their sum vanishes $\sum_{(i)} k_{(i)} = 0$ \cite{Jafferis:2008qz}. The $n$ edges are also indexed by a $(j)$ which takes values from $1$ to $n$; the $(j)$'th edge is the one that joins the $(j)$'th node to the $(j+1)$'th node and we should take $(n+1)$'th edge to be $(1)$'st edge.  At each edge there is a $\mathcal{N}=3$ hyper multiplet which consists of two $\mathcal{N}=2$ chiral superfields $(\mathcal{Z}_{(j)}, \mathcal{W}_{{(j)}})$.  $\mathcal{Z}_{(j)}$ is a bifundamental field transforming in the fundamental representation of $U(N)_{(j)}$ and in the anti-fundamental representation of $U(N)_{(j+1)}$ while $\mathcal{W}_{(j)}$ transforms in the anti-fundamental representation of $U(N)_{(j)}$ and in the fundamental representation of $U(N)_{(j+1)}$. A special case of this class of theories is the ABJM theory \cite{Aharony:2008ug} for which the number of nodes is $n=2$. Our superfield and other conventions follow \cite{Benna:2009xd} whose work on ABJM is being followed here in this paper for Jafferis-Tomasiello theories. These theories are $\mathcal{N}=3$ SCFT's following  \cite{Gaiotto:2007qi} and in this paper we begin a study of monopole operators along the lines of Benna, Klebanov and Klose (BKK) in \cite{Benna:2009xd}.

A crucial ingredient in the BKK \cite{Benna:2009xd} set up is the UV completion, which is obtained by adding Yang-Mills terms to the gauge fields and thus introduces the dimensionful Yang-Mills coupling $g$. They also add dynamical fields in the adjoint representation in order to preserve $\mathcal{N}=3$ supersymmetry and $SU(2)_R$ symmetry. Thus the UV completion is a $\mathcal{N}=3$ Chern-Simons Yang-Mills non-conformal theory. The Yang-Mills term is irrelevant in three dimensions and there is a renormalisation group flow from the $\mathcal{N}=3$ Chern-Simons-Yang-Mills theory in the UV ($g=0$) to a conformal fixed point in the IR ($g=\infty$). In the infra red, the Yang-Mills terms and the kinetic terms for the adjoint matter fields go to zero and the equations of motion become constraints allowing one to integrate them out and recover the superconformal field theory.  The flow itself preserves $\mathcal{N}=3$ supersymmetry and $SU(2)_R$ symmetry. There are other UV completions available in the literature for the superconformal field theories under study such as the one used in \cite{Kim:2009wb} and in  \cite{Bashkirov:2010kz}, where a smaller amount of supersymmetry is preserved along the flow. 

Our aim in this work is to study monopole operators in the Jafferis-Tomasiello class of superconformal field theories. Monopole operators \cite{Borokhov:2002ib, Borokhov:2002cg, Borokhov:2003yu} are very important objects and subject of a wide range of studies \cite{Pufu:2013eda, Dyer:2015zha, Dyer:2013fja, Intriligator:2014fda} contributing to such interesting phenomena as supersymmetry enhancement in ABJM at low levels \cite{Aharony:2008ug, Benna:2009xd, Bashkirov:2010kz} and to matching of the spectra in AdS-CFT studies. Monopole operators are defined as objects which when inserted at a point create quantized flux in a $U(1)$ subgroup of the gauge group through a two sphere surrounding the point.  The flux inserted by a monopole operator has the following forms 
\begin{equation}\label{1.1}
A|_{R^3} = \frac{\mathbf{H}}{2} \frac{\pm 1 - \cos \theta}{r}\,d\varphi, \qquad A|_{R \times S^2} = \frac{\mathbf{H}}{2} (\pm 1 - \cos \theta)\,d\varphi
\end{equation}
for the gauge theory defined on $\mathbf{R}^3$ and in the radial quantization picture on $\mathbf{R} \times \mathbf{S}^2$ respectively.
Supersymmetric monopole operators when inserted at a point  will have other fields also turned on other than the gauge field. In Chern-Simons gauge theories with $U(N)$ gauge group, monopole operators are characterised by a generator $\mathbf{H}$ that specifies the embedding of $U(1)$ into $U(N)$. The gauge field is proportional to this generator $\mathbf{H} = \mathop{\mathrm{diag}} \{q_1, q_2 \ldots q_N\}$ where the $q_i$'s are the magnetic charges. Monopole operators also transform in specific  gauge representations given by the data of the Chern-Simons level and the magnetic charges \cite{Kapustin:2006pk}, see also \cite{Klebanov:2008vq}. Monopole operators in ABJM theories are specified by two sets of magnetic charges $\mathbf{H}_1$ and $\mathbf{H}_2$ constrained to be such that $\mathop{\mathrm{tr}}\mathbf{H}_1 = \mathop{\mathrm{tr}}\mathbf{H}_2$ \cite{Kim:2009wb}. BKK in \cite{Benna:2009xd} study monopole operators with $\mathbf{H}_1 = \mathbf{H}_2$ and they find even within this class of operators the interesting ones needed for supersymmetry enhancement and matching with supergravity spectra. The generic monopole operator in Jafferis-Tomasiello theory must be specified by $n$ sets of magnetic charges $\mathbf{H}_{(j)}$ but we will only investigate a subclass 
with all of them identical i.e $\mathbf{H}_{(1)} = \mathbf{H}_{(2)} = \ldots = \mathbf{H}_{(n)}\equiv \mathbf{H}$. For the purposes of this paper, we do not need to discuss the gauge representations of monopole operators in Jafferis-Tomasiello theories; we will do so in a future work \cite{nextwork} where we also compute the charges and scaling dimensions of these monopole operators along the lines of \cite{Benna:2009xd}.

Supersymmetric monopole operators in the superconformal field theory at the IR fixed point are studied via classical BPS monopole solutions of the UV Chern-Simons-Yang-Mills theory. The way this comes about is explained nicely in \cite{Benna:2009xd}, where they recognise that a ``separation of scales exist between the BPS background that inserts a flux at a space-time point in accordance with \eqref{1.1}  - recall that its magnitude is constrained by the Dirac quantisation condition - and the typical size of quantum fluctuation of fields.'' Therefore, they \cite{Benna:2009xd} say, one can ``treat the monopole operator as a classical background.'' Hence, in this paper, to study supersymmetric monopole operators of Jafferis-Tomasiello superconformal Chern-Simons theory, we will obtain the classical BPS monopole solutions to the $\mathcal{N}=3$ Yang-Mills deformation of Jafferis-Tomasiello theory. 

The paper is organised as follows. In the following section \ref{2}, first we give the superspace action for the $\mathcal{N}=3$ Yang-Mills deformation of the Jafferis-Tomasiello theory in $\mathbf{R}^{1,2}$ in \ref{211}. Then in \ref{212} we give the component action written in terms of fields arranged into $SU(2)_R$ multiplets. Such a component action with manifest $SU(2)_R$ symmetry is needed for the $SU(2)_R$ charge computation which we will take up in \cite{nextwork}. After this in \ref{213} we give the equations for the supersymmetry transformation of all fields.  Then in \ref{214} we perform a rescaling on the fermions in the $\mathcal{N}=3$ vector multiplets which is the scaling needed to consider them as quantum fluctuations in the UV (see section 3.1 of \cite{Benna:2009xd}). This rescaling affects the component action as well as the supersymmetry transformations which we make explicit. In the latter part of the second section \ref{22}, we make the transition from $\mathbf{R}^{1,2}$ to $\mathbf{R}\times S^2$ and write out the component action \ref{221} and the supersymmetry transformations \ref{222}. After all this, in section \ref{3}, we study the vanishing conditions on the supersymmetry transformations of the fermions to obtain classical BPS monopole solutions. We study in detail the three node quiver in \ref{31} and from the lessons learnt in that special case, we give the solutions for the general $n$-node quiver in \ref{32}. We end with a concluding section \ref{4} on possible future directions (including \cite{nextwork}.) An appendix \ref{A} consists of conventions.

\section{\label{2} The action and supersymmetries of the $\mathcal{N}=3$ Yang-Mills deformed Jafferis-Tomasiello theory.}

In this section we first construct the action and the supersymmetries of the $\mathcal{N}=3$ Chern-Simons Yang-Mills theory first in $\mathbf{R^{1,2}}$ and then in $\mathbf{R}\times \mathbf{S}^2$.

\subsection{\label{21}$\mathbf{R^{1,2}}$}

\subsubsection{\label{211}Superspace action}

At each node we have a $\mathcal{N} = 3$ vector multiplet\footnote{\label{note1}This is really a $\mathcal{N} = 4$ multiplet. When the theory includes Chern-Simons terms, the supersymmetry is reduced to $\mathcal{N} = 3$.} which is comprised of the $\mathcal{N} = 2$ super fields $(\mathcal{V}_{(j)}, \Phi_{{(j)}})$ in the adjoint representation of $U(N)_{(j)}$.  For each edge, we have a $\mathcal{N} = 3$ hyper multiplet\footref{note1} which is comprised of two $\mathcal{N} = 2$ chiral super fields $(\mathcal{Z}_{(j)}, \mathcal{W}_{{(j)}})$ in bifundamental representations.  The component fields for each of the superfields can be read off from appendix \ref{A} which contains our notation and conventions and which closely follow \cite{Benna:2009xd}.

The $\mathcal{N}=3$ action on $\mathbf{R^{1,2}}$ consists of five parts. The first three involve the vector multiplet fields and the last two involve the hyper multiplet fields together with their minimal couplings to the vector multiplet fields. 
\begin{eqnarray} \label{2.2}
\mathcal{S} = \mathcal{S}_{\mathrm{CS}} + \mathcal{S}_{\mathrm{YM}} + \mathcal{S}_{\mathrm{adj}} + \mathcal{S}_{\mathrm{mat}} + \mathcal{S}_{\mathrm{pot}}
\end{eqnarray}
The first of the five parts is the sum over the nodes of the Chern-Simons terms. The levels satisfy \cite{Jafferis:2008qz} the condition $\sum_{(j)=1}^n k_{(j)} = 0$.
\begin{eqnarray} \label{CS}
\mathcal{S}_{\mathrm{CS}}=-\frac{i}{8\pi}\int d^3x ~ d^4\theta \int_0^1 ds\sum_{(j)=1}^n \mathop{\mathrm{tr}}\Big[k_{(j)} \mathcal{V}_{(j)}\overline{D}^\alpha(e^{s\mathcal{V}_{(j)}} D_\alpha e^{-s\mathcal{V}_{{(j)}}})\Big]
\end{eqnarray}
The $s$ integral is the convenient way of writing the non-Abelian Chern-Simons action. The second of the five parts consists of the Yang-Mills terms, one for each node of the quiver. 
\begin{eqnarray} \label{YM}
  \mathcal{S}_{\mathrm{YM}} = \frac{1}{4g^2} \int d^3x ~ d^2\theta  \sum_{(j)=1}^n \mathop{\mathrm{tr}}  \Big[{\mathcal{U}^{\alpha}}_{(j)} {\mathcal{U}_{\alpha}}_{(j)} \Big]  \; 
\end{eqnarray}
$g$ is the dimensionful coupling (of mass dimension $\frac12$) responsible for the RG flow. ${\mathcal{U}^{\alpha}}_{(j)}$ is the field strength chiral superfield given by ${\mathcal{U}^{\alpha}}_{(j)} = \frac14 \overline{D}^2\,e^{\mathcal{V}_{{(j)}}}\,D_\alpha\,e^{-\mathcal{V}_{{(j)}}}.$
The third of the five parts makes the other adjoint fields in the $\mathcal{N}=3$ vector multiplet dynamical. 
\begin{eqnarray} \label{adj}
  \mathcal{S}_{\mathrm{adj}} = \frac{1}{g^2} \int d^3x~d^4\theta\: \sum_{{{(j)}}=1}^n\mathop{\mathrm{tr}} \Big[ - \overline{\Phi}_{{(j)}} e^{-\mathcal{V}_{{(j)}}} \Phi_{{(j)}} e^{\mathcal{V}_{{(j)}}} \Big] \; 
\end{eqnarray}
The fourth of the five parts is the minimally coupled action for the bifundamental fields in the hyper multiplets and is a sum over the edges of the quiver. 
\begin{eqnarray} \label{mat}
\mathcal{S}_{\mathrm{mat}}=\int d^3x ~ d^4\theta \sum_{{{(j)}}=1}^n\mathop{\mathrm{tr}}\Big[- \overline{\mathcal{Z}}_{{(j)}} e^{-\mathcal{V}_{{(j)}}}\mathcal{Z}_{{(j)}} e^{\mathcal{V}_{{(j+1)}}}-\overline{\mathcal{W}}_{{(j)}} e^{-\mathcal{V}_{{(j+1)}}}\mathcal{W}_{{(j)}} e^{\mathcal{V}_{{(j)}}}\Big]
\end{eqnarray}
The fifth of the five parts is the super potential term
\begin{eqnarray} \label{pot}
\mathcal{S}_{\mathrm{pot}}=\int d^3x ~ d^2\theta \sum_{{{(j)}}=1}^n W_{{(j)}}-\int d^3x ~ d^2\overline{\theta} \sum_{{{(j)}}=1}^n \overline{W}_{{(j)}}
\end{eqnarray}
where
\begin{eqnarray} \label{superpot}
W_{{(j)}}=\mathop{\mathrm{tr}} \Big[\Phi_{{(j)}} \mathcal{Z}_{{(j)}} \mathcal{W}_{{(j)}} - \Phi_{{(j)}} \mathcal{W}_{{(j-1)}}\mathcal{Z}_{{(j-1)}})+\frac{k_{{(j)}}}{8\pi}\mathop{\mathrm{tr}}(\Phi_{{(j)}} \Phi_{{(j)}} \Big].
\end{eqnarray}
We can recover the Jafferis-Tomasiello super conformal field theory from the above as follows. At the infra red fixed point $g \rightarrow \infty$ which sets $\mathcal{S}_{\mathrm{YM}}$ and $\mathcal{S}_{\mathrm{adj}}$ to zero thus making the gauge fields and the adjoint fields non-dynamical. We can then integrate out the $\Phi_{{(j)}}$'s and recover the Jafferis-Tomasiello theory.

\subsubsection{\label{212}Component action in terms of $SU(2)_R$ multiplets}

From the superspace action \eqref{2.2}, we obtain the component action by doing the Grassmann integrals and integrating out the auxiliary fields $D_{(j)}$, $F_{\phi\,(j)}$, $F_{(j)}$ and the $G_{(j)}$'s. This is very big expression and we won't give it here.

$SU(2)_R$ symmetry is preserved along the RG flow. In fact, the computation of the scaling dimensions of supersymmetric monopole operators in the ABJM case was achieved in \cite{Benna:2009xd} by computing the $SU(2)_R$ charges in the UV, which being unchanging along the flow, gets related to the scaling dimension in the IR by supersymmetry.  Hence it is desirable to write the component action in which the $SU(2)_R$ symmetry is manifest.  Following \cite{Benna:2009xd}, we organise the non-auxiliary component fields into following four $SU(2)_R$ multiplets. 

The first $SU(2)_R$ multiplet consists of the adjoint scalar fields in the vector multiplet which 
transform in the $\mathbf{3}$: 
\[{\phi^{~a}_b}_{{(j)}} = \phi_{{(j)}\,i} ~(\sigma_i)^{~a}_b=\left( \begin{array}{cc}
 -\sigma_{{(j)}} & \phi_{{(j)}}^\dagger \\
     \phi_{{(j)}} & \sigma_{{(j)}}
\end{array} \right).\] 
${\phi^{~a}_b}_{{(j)}}$'s are traceless hermitian matrices which is expressed by the equation:
\begin{equation}
({\phi^{~a}_b}_{{(j)}})^* = {\phi^{~b}_a}_{{(j)}} = \epsilon_{ac} \, \epsilon^{bd} \,{\phi^{~c}_d}_{{(j)}}.
\end{equation}
The second $SU(2)_R$ multiplet consists of the fermions in the vector multiplet which 
transform in the reducible representation $\mathbf{2}\times\mathbf{2}$: 
\begin{equation}
\lambda^{ab}_{{(j)}}=\left( \begin{array}{cc}
 {\chi_{\sigma}}_{{(j)}} e^{-i\pi/4} & {\chi_{\phi}^\dagger}_{{(j)}} \, e^{-i\pi/4} \\
    {\chi_{\phi}}_{{(j)}}  e^{+i\pi/4} & -{\chi_{\sigma}^\dagger}_{{(j)}} \, e^{+i\pi/4}
\end{array}  \right)
\end{equation}
The ${\lambda^{ab}}_{{(j)}}$'s s are generic matrices with the following properties:
\begin{eqnarray}
(\lambda^{ab}_{{(j)}})^* = -{\lambda_{ab}}_{{(j)}} = -\epsilon_{ac} \epsilon_{bd} \lambda^{cd}_{{(j)}}
\end{eqnarray}
One can extract the irreducible components $\mathbf{1}$ and $\mathbf{3}$ by computing ${\lambda^{ab}}_{{(j)}}\,\epsilon_{ba}$ and ${\lambda^{ab}}_{{(j)}}\,(\sigma_i)_{ba}$ respectively. 
The third $SU(2)_R$ multiplet consists of the scalar fields in the hyper multiplet which transform as a doublet:
\begin{eqnarray}\label{eqn:X-SU2R} 
X^{a}_{{(j)}} =\left( \begin{array}{cc}
Z_{{(j)}} \\ W^{\dagger }_{{(j)}}
\end{array}\right),\hspace{1cm}
{X^\dagger_{a}}_{{(j)}} = \left( \begin{array}{cc} 
Z^\dagger_{{(j)}} \\ W_{{(j)}}.
\end{array} \right)
\end{eqnarray}
The last $SU(2)_R$ multiplet consists of the fermions in the hyper multiplet which transform as a doublet:
\begin{equation}
\label{eqn:xi-SU2R}
\xi^{a}_{{(j)}} =\left( \begin{array}{cc}
\omega^\dagger_{{(j)}}  \, e^{i\pi/4} \\ \zeta_{{(j)}} \, e^{-i\pi/4} 
\end{array}\right) ,\hspace{1cm}
{\xi^\dagger_{a}}_{{(j)}} =\left(\begin{array}{cc}  \omega_{{(j)}} \, e^{-i\pi/4} \\ \zeta^\dagger_{{(j)}} \, e^{i\pi/4} 
\end{array}\right).
\end{equation}

After a long computation, we can show that the component action  written in terms of the above $SU(2)_R$ multiplets is:
\begin{multline} \label{Skin}
\mathcal{S}_\mathrm{kin} = \int d^3x \sum_{{{(j)}}=1}^n \mathop{\mathrm{tr}} \Big[- \frac{1}{2g^2} F^{\mu\nu}_{{(j)}} {F_{\mu\nu}}_{{(j)}}+ \kappa_{{(j)}}  \epsilon^{\mu\nu\lambda} \big({ {A_\mu}_{{(j)}} \partial_\nu {A_\lambda}_{{(j)}} + \frac{2i}{3} {A_\mu}_{{(j)}} {A_\nu}_{{(j)}} {A_\lambda}_{{(j)}} }\big)  \\ \hspace{17mm}
   - \mathcal{D}_\mu X^\dagger_{{(j)}} \mathcal{D}^\mu X_{{(j)}}
   + i \,\xi^\dagger_{{(j)}} \slashed{\mathcal{D}} \xi_{{(j)}}
   \\[1mm]\hspace{17mm}
   - \frac{1}{2g^2} \mathcal{D}_\mu {\phi^a_b}_{{(j)}} \mathcal{D}^\mu {\phi^b_a}_{{(j)}}
   - \frac{1}{2} \kappa_{{(j)}}^2 g^2 \, {\phi^a_b}_{{(j)}} {\phi^b_a}_{{(j)}}
      \\[1mm]\hspace{17mm}
   - \frac{i}{2g^2} {\lambda^{ab}}_{{(j)}} \slashed{\mathcal{D}} {\lambda_{ab}}_{{(j)}}
   - \frac{\kappa_{{(j)}}}{2} \, i {\lambda^{ab}}_{{(j)}} {\lambda_{ba}}_{{(j)}}
   \Big]
\end{multline}
and
\begin{eqnarray} \label{Sint}
\mathcal{S}_{\mathrm{int}} = \int d^3x \sum_{(j)=1}^n \mathop{\mathrm{tr}} \Big[- \kappa_{{(j)}} g^2 \, {X^\dagger_a}_{{(j)}} {\phi^a_b}_{{(j)}} X^b_{{(j)}} + \kappa_{{(j)}} g^2 \, X^a_{{(j-1)}} {{\phi}^b_a}_{{(j)}} {X^\dagger_b}_{{(j-1)}} - i {\xi^\dagger_a}_{{(j)}} {\phi^a_b}_{{(j)}} \xi^b_{{(j)}} \nonumber \\+ i {\xi^a}_{{(j-1)}} {\phi^b_a}_{{(j)}} {\xi^\dagger_b}_{{(j-1)}}  + \epsilon_{ac} \lambda^{cb}_{{(j)}} X^a_{{(j)}} {\xi^\dagger_b}_{{(j)}}- \epsilon^{ac} {\lambda_{cb}}_{{(j)}} \xi^b_{{(j)}} {X^\dagger_a}_{{(j)}}-\epsilon_{ac} {\lambda}^{cb}_{{(j)}} {\xi^\dagger_b}_{{(j-1)}} X^a_{{(j-1)}} \nonumber \\+\epsilon^{ac} {\lambda_{cb}}_{{(j)}} {X^\dagger_a}_{{(j-1)}} \xi^b_{{(j-1)}}  {{-}} \frac{\kappa_{{(j)}}}{6} {\phi^a_b}_{{(j)}} [{\phi^b_c}_{{(j)}}},{{\phi^c_a}_{{(j)}}]        - \frac{1}{2g^2} i {\lambda_{ab}}_{{(j)}}  [{\phi^b_c}_{{(j)}}},{{\lambda^{ac}}_{{(j)}}]    \nonumber \\[1mm]\hspace{17mm} - \frac{g^2}{4} (X_{{(j)}} \sigma_i X^\dagger_{{(j)}}) (X_{{(j)}} \sigma_i X^\dagger_{{(j)}})- \frac{g^2}{4} (X^\dagger_{{(j)}} \sigma_i X_{{(j)}}) (X^\dagger_{{(j)}} \sigma_i X_{{(j)}}) \nonumber \\ + \frac{g^2}{2} \big((X_{{(j)}} \sigma_i X^\dagger_{{(j)}}) (X^\dagger_{{(j-1)}} \sigma_i X_{{(j-1)}})\big) \nonumber \\[1mm]\hspace{17mm}  - \frac{1}{2} (X_{{(j)}} X^\dagger_{{(j)}}) {\phi^a_b}_{{(j)}} {\phi^b_a}_{{(j)}} - \frac{1}{2} (X^\dagger_{{(j-1)}} X_{{(j-1)}}) {\phi^a_b}_{{(j)}} {\phi^b_a}_{{(j)}} + {X^\dagger_a}_{{(j)}} {\phi^b_c}_{{(j)}} X^{a}_{{(j)}} {\phi^c_b}_{{(j+1)}} \nonumber \\[1mm]\hspace{17mm}+ \frac{1}{8g^2} [{\phi^a_b}_{{(j)}}},{{\phi^c_d}_{{(j)}}] [{\phi^b_a}_{{(j)}}},{{\phi^d_c}_{{(j)}}]  \Big]
\end{eqnarray}
where $\kappa_{{(j)}} \equiv \frac{k_{{(j)}}}{4\pi}$,  $X_{(j)} \sigma_i X^\dagger_{(j)} \equiv X^{a}_{(j)} (\sigma_i)_a{}^b {X^\dagger_{b}}_{(j)}$ and $X^\dagger_{(j)} \sigma_i X_{(j)} \equiv {X^\dagger_{a}}_{(j)} (\sigma_i)^a{}_b X^{b}_{(j)}$. The $(\sigma_i)_a{}^b = \sigma_i$ are the usual Pauli matrices and the $(\sigma_i)^a{}_b = \sigma_i^{\scriptscriptstyle\mathrm{T}}$ are the transpose of the Pauli matrices. The various gauge covariant derivatives above are
\begin{multline} 
{F_{\mu\nu}}_{(j)}=\partial_\mu {A_\nu}_{(j)}-\partial_\nu {A_\mu}_{(j)}+i[{A_\mu}_{(j)},{A_\nu}_{(j)}] 
\nonumber \\
\mathcal{D}_\mu {\phi^{~a}_b}_{{(j)}} = \partial_\mu{\phi^{~a}_b}_{{(j)}} + i [{A_\mu}_{(j)},{\phi^{~a}_b}_{{(j)}}], \qquad
\mathcal{D}_\mu{\lambda^{ab}}_{{(j)}} = \partial_\mu{\lambda^{ab}}_{{(j)}}+i[{A_\mu}_{(j)}, {\lambda^{ab}}_{{(j)}}] 
\nonumber \\
\mathcal{D}_\mu X^a_{{(j)}}=\partial_\mu X^a_{{(j)}} +i{A_\mu}_{{(j)}} X^a_{{(j)}}-iX^a_{{(j)}} {A_\mu}_{{(j+1)}}, \qquad
\mathcal{D}_\mu \xi^\dagger_{a {(j)}}=\partial_\mu \xi^\dagger_{a {(j)}} +i{A_\mu}_{{(j)}} \xi^\dagger_{a {(j)}}-i \xi^\dagger_{a {(j)}} {A_\mu}_{{(j+1)}}. \nonumber
\end{multline}

\subsubsection{\label{213}Supersymmetry transformations}

The supersymmetry transformation parameters in  a $\mathcal{N}=3$ theory are in the  $\mathbf{3}$ of $SU(2)_R$: $\quad \varepsilon_{ab} = \varepsilon_i\,(\sigma_i)_{ab}$. The supersymmetry transformations of the non-auxiliary component fields that leave the action \eqref{Skin}, \eqref{Sint} invariant (after a long computation) are 
\begin{eqnarray}
\delta A_{\mu\,(j)} = -\frac{i}{2}\,\varepsilon_{ab}\,\gamma_\mu\,\lambda^{ab}_{\,(j)}  \\
\delta \lambda^{ab}_{\,(j)} = \frac12 \epsilon^{\mu\nu\lambda}\,F_{\mu \nu\,(j)} \gamma_\lambda\,\varepsilon^{ab} - i \slashed{\mathcal{D}} \phi^b_{c\,(j)}\varepsilon^{ac} + \frac{i}{2} [{\phi^b_c}_{{(j)}}},{{\phi^c_d}_{{(j)}}] \varepsilon^{ad} + \kappa_{(j)} g^2 i \phi^b_{c\,(j)}\varepsilon^{ac} \nonumber \\  + g^2 i (X^a_{(j)}\,X^\dagger_{c\,(j)} \varepsilon^{cb} - \varepsilon^{bc} X^\dagger_{c\,(j-1)}\,X^a_{(j-1)}) - \frac{i g^2}{2} (X_{(j)}\,X^\dagger_{(j)} - X^\dagger_{(j-1)}X_{(j-1)} ) \varepsilon^{ab} \\
\delta \phi^a_{b\,(j)} = - \varepsilon_{cb} \lambda^{ca}_{\,(j)} + \frac12 \delta^a_{b} \varepsilon_{cd}\,\lambda^{cd}_{\,(j)} \\
\delta X^a_{(j)} = - i \,\varepsilon^a_b\,\xi^b_{(j)}, \qquad \delta \xi^a_{(j)} =  \slashed{\mathcal{D}} X^b_{(j)} \varepsilon^a_b  +  \phi^a_{b\,(j)} \varepsilon^b_c\, X^c_{(j)} - X^c_{(j)}\varepsilon^b_c\, \phi^a_{b\,(j+1)}  \\
\delta X^\dagger_{a\,(j)} = -i \,\xi^\dagger_{b\,(j)} \varepsilon^b_a, \qquad  \delta \xi^\dagger_a = \slashed{\mathcal{D}} X^\dagger_{b\,(j)} \varepsilon^b_a  - \phi^b_{a\,(j+1)} \varepsilon^c_b \,X^\dagger_{c\,(j)} +  X^\dagger_{c\,(j)}  \varepsilon^c_b\, \phi^b_{a\,(j)}.
\end{eqnarray}

\subsubsection{\label{214}Scaling the vector multiplet fermions}
Following BKK \cite{Benna:2009xd}, monopole operators in the IR theory correspond to classical bosonic BPS monopole solutions of the UV theory. The fermions in the vector multiplet are to be treated as fluctuations in the UV. Towards this end, we need to appropriately scale the vector multiplet fermions as follows:
\begin{equation}
\lambda = g\, \lambda'
\end{equation}
This changes some terms in the action. The last line in \eqref{Skin} becomes
$   - \frac{i}{2} {\lambda'^{ab}}_{{(j)}} \slashed{\mathcal{D}} {\lambda'_{ab}}_{{(j)}}
   - \frac{\kappa_{{(j)}}g^2}{2} \, i {\lambda'^{ab}}_{{(j)}} {\lambda'_{ba}}_{{(j)}} $ and the second and third lines of \eqref{Sint} are similarly modified. More importantly for us, the supersymmetry transformations are modified as follows:
\begin{eqnarray}
\delta A_{\mu\,(j)} = -\frac{i g}{2}\,\varepsilon_{ab}\,\gamma_\mu\,\lambda'^{ab}_{\,(j)} \\
\delta \lambda'^{ab}_{\,(j)} = \frac{1}{2g} \epsilon^{\mu\nu\lambda}\,F_{\mu \nu\,(j)} \gamma_\lambda\,\varepsilon^{ab} - \frac{i}{g} \slashed{\mathcal{D}} \phi^b_{c\,(j)}\varepsilon^{ac} + \frac{i}{2g} [{\phi^b_c}_{{(j)}}},{{\phi^c_d}_{{(j)}}] \varepsilon^{ad} + \kappa_{(j)}  i g \,\phi^b_{c\,(j)}\varepsilon^{ac} \nonumber \\  +  i g (X^a_{(j)}\,X^\dagger_{c\,(j)} \varepsilon^{cb} - \varepsilon^{bc} X^\dagger_{c\,(j-1)}\,X^a_{(j-1)}) - \frac{i g}{2} (X_{(j)}\,X^\dagger_{(j)} - X^\dagger_{(j-1)}X_{(j-1)} ) \varepsilon^{ab}\\
\delta \phi^a_{b\,(j)} = - g \varepsilon_{cb} \lambda'^{ca}_{\,(j)} + \frac{g}{2} \delta^a_{b} \varepsilon_{cd}\,\lambda'^{cd}_{\,(j)}
\end{eqnarray}

\subsection{\label{22}$\mathbf{R} \times \mathbf{S^2}$}

The goal is to compute the scaling dimensions of supersymmetric monopole operators, which is achieved by working in the radial quantisation picture. Hence we need to define the super conformal field theory as well as the entire RG flow on $\mathbf{R}\times \mathbf{S}^2$. We again follow the conventions of BKK \cite{Benna:2009xd}. First we go from $\mathbf{R}^{1,2}$ to $\mathbf{R}^3$ by defining Euclidean co-ordinates $\{x^1, x^2, x^3\} = \{x^1, x^2, i x^0\}$. Then we change to polar co-ordinates $\{r, \theta, \phi\}$ after which we we introduce the new variable $\tau$ by $r = e^\tau$. The result is a theory on $\mathbf{R}\times \mathbf{S}^2$ with co-ordinates $\{ \tau, \theta, \phi\}$ and metric $ ds^2 = d\tau^2 + d\theta^2 + \sin^2\theta \,d\phi^2$.

\subsubsection{\label{221}Component action in terms of $SU(2)_R$ multiplets}
One needs to rescale the $\mathbf{R}^{1,3}$ fields in the following way to obtain the $\mathbf{R}\times \mathbf{S}^2$ fields, denoted below with the tilde $\sim$ symbol.
\begin{eqnarray}{A_\mu}_{{(j)}} = e^{-\tau} \tilde{A_\mu}_{{(j)}}, \qquad {\phi^a_b}_{{(j)}}  = e^{-\tau} \tilde{\phi^a_b}_{{(j)}}, \qquad \lambda'^{ab}_{{(j)}} = e^{-\tau} \tilde{\lambda'^{ab}_{{(j)}}}, \nonumber \\  X^{a}_{{(j)}} = e^{-\frac{\tau}{2}} \tilde{X}^{a}_{{(j)}} \qquad \xi^{a}_{{(j)}} = e^{-\tau}  \tilde{\xi}^{a}_{{(j)}},\qquad g = e^{-\frac{\tau}{2}} \tilde{g}
\end{eqnarray}
One obtains the following component action, wherein we have dropped the tildes (except for $\tilde{g}$) and the primes:
\begin{multline} \label{SkinS}
\mathcal{S}^E_\mathrm{kin} = \int d\tau d\Omega \sum_{{{(j)}}=1}^n \mathop{\mathrm{tr}} \Big[\frac{1}{2\tilde{g}^2} F^{mn}_{{(j)}} {F_{mn}}_{{(j)}} - \kappa_{{(j)}}  i \epsilon^{mnk} \big({ {A_m}_{{(j)}} \partial_n {A_k}_{{(j)}} + \frac{2i}{3} {A_m}_{{(j)}} {A_n}_{{(j)}} {A_k}_{{(j)}} }\big)  \\ \hspace{17mm}
   + \mathcal{D}_\mu X^\dagger_{{(j)}} \mathcal{D}^\mu X_{{(j)}} + \frac14 X^\dagger_{{(j)}} X_{{(j)}}
   - i \,\xi^\dagger_{{(j)}} \slashed{\mathcal{D}} \xi_{{(j)}} \\
   + \frac{1}{2\tilde{g}^2} \mathcal{D}_\mu {\phi^a_b}_{{(j)}} \mathcal{D}^\mu {\phi^b_a}_{{(j)}}
   + \frac{1}{2} \kappa_{{(j)}}^2 \tilde{g}^2 \, {\phi^a_b}_{{(j)}} {\phi^b_a}_{{(j)}}
      \\
   + \frac{i}{2} {\lambda^{ab}}_{{(j)}} \slashed{\mathcal{D}} {\lambda_{ab}}_{{(j)}}
   + \frac{\kappa_{{(j)}}\tilde{g}^2}{2} \, i {\lambda^{ab}}_{{(j)}} {\lambda_{ba}}_{{(j)}}
   \Big]
\end{multline}
and
\begin{eqnarray} \label{SintS}
\mathcal{S}^E_{\mathrm{int}} = \int d\tau d\Omega \sum_{(j)=1}^n \mathop{\mathrm{tr}} \Big[ \kappa_{{(j)}} \tilde{g}^2 \, {X^\dagger_a}_{{(j)}} {\phi^a_b}_{{(j)}} X^b_{{(j)}} - \kappa_{{(j)}} \tilde{g}^2 \, X^a_{{(j-1)}} {{\phi}^b_a}_{{(j)}} {X^\dagger_b}_{{(j-1)}} + i {\xi^\dagger_a}_{{(j)}} {\phi^a_b}_{{(j)}} \xi^b_{{(j)}} \nonumber \\- i {\xi^a}_{{(j-1)}} {\phi^b_a}_{{(j)}} {\xi^\dagger_b}_{{(j-1)}}  - \tilde{g}\epsilon_{ac} \lambda^{cb}_{{(j)}} X^a_{{(j)}} {\xi^\dagger_b}_{{(j)}} + \tilde{g} \epsilon^{ac} {\lambda_{cb}}_{{(j)}} \xi^b_{{(j)}} {X^\dagger_a}_{{(j)}} + \tilde{g} \epsilon_{ac} {\lambda}^{cb}_{{(j)}} {\xi^\dagger_b}_{{(j-1)}} X^a_{{(j-1)}} \nonumber \\- \tilde{g}\epsilon^{ac} {\lambda_{cb}}_{{(j)}} {X^\dagger_a}_{{(j-1)}} \xi^b_{{(j-1)}}  {{-}} \frac{\kappa_{{(j)}}}{6} {\phi^a_b}_{{(j)}} [{\phi^b_c}_{{(j)}}},{{\phi^c_a}_{{(j)}}]        + \frac{i}{2}  {\lambda_{ab}}_{{(j)}}  [{\phi^b_c}_{{(j)}}},{{\lambda^{ac}}_{{(j)}}]    \nonumber \\[1mm]\hspace{17mm} + \frac{\tilde{g}^2}{4} (X_{{(j)}} \sigma_i X^\dagger_{{(j)}}) (X_{{(j)}} \sigma_i X^\dagger_{{(j)}}) + \frac{\tilde{g}^2}{4} (X^\dagger_{{(j)}} \sigma_i X_{{(j)}}) (X^\dagger_{{(j)}} \sigma_i X_{{(j)}}) \nonumber \\ -\frac{\tilde{g}^2}{2} \big((X_{{(j)}} \sigma_i X^\dagger_{{(j)}}) (X^\dagger_{{(j-1)}} \sigma_i X_{{(j-1)}})\big) \nonumber \\[1mm]\hspace{17mm} + \frac{1}{2} (X_{{(j)}} X^\dagger_{{(j)}}) {\phi^a_b}_{{(j)}} {\phi^b_a}_{{(j)}} + \frac{1}{2} (X^\dagger_{{(j-1)}} X_{{(j-1)}}) {\phi^a_b}_{{(j)}} {\phi^b_a}_{{(j)}} - {X^\dagger_a}_{{(j)}} {\phi^b_c}_{{(j)}} X^{a}_{{(j)}} {\phi^c_b}_{{(j+1)}} \nonumber \\[1mm]\hspace{17mm}- \frac{1}{8\tilde{g}^2} [{\phi^a_b}_{{(j)}}},{{\phi^c_d}_{{(j)}}] [{\phi^b_a}_{{(j)}}},{{\phi^d_c}_{{(j)}}]  \Big].
\end{eqnarray}

\subsubsection{\label{222}Supersymmetry transformations}
We need the supersymmetry transformations in $\mathbf{R}\times\mathbf{S}^2$. The supersymmetry transformation parameter is rescaled as follows \cite{Benna:2009xd}
\begin{equation}
 \tilde{\varepsilon}_{ab}(\tau) = e^{-\frac\tau2}\,\varepsilon_{ab} 
 \end{equation}
and it solves the Killing spinor equation on  $\mathbf{R}\times\mathbf{S}^2$
 \begin{equation}\label{ksp}
 \nabla_m\,\tilde{\varepsilon}_{ab} = - \frac12 \, \gamma_m \,\gamma^\tau \tilde{\varepsilon}_{ab}.
  \end{equation}
The supersymmetry transformations that leave the action \eqref{SkinS}, \eqref{SintS} invariant are,  after dropping the tildes on the parameters:
\begin{eqnarray} 
\delta A_{m\,(j)} = -\frac{i \tilde{g}}{2}\,\varepsilon_{ab}\,\gamma_m\,\lambda^{ab}_{\,(j)} \\ \label{sp-susy}
\delta \lambda^{ab}_{\,(j)} = \frac{i}{2\tilde{g}} \epsilon^{m n k}\,F_{m n \,(j)} \gamma_k\,\varepsilon^{ab} - \frac{i}{\tilde{g}} \slashed{\mathcal{D}} \phi^b_{c\,(j)}\varepsilon^{ac} - \frac{2 i}{3 \tilde{g}} \phi^b_{c\,(j)} \slashed{\nabla} \varepsilon^{ac}+ \frac{i}{2\tilde{g}} [{\phi^b_c}_{{(j)}}},{{\phi^c_d}_{{(j)}}] \varepsilon^{ad} \nonumber \\ + \kappa_{(j)}  i \tilde{g} \,\phi^b_{c\,(j)}\varepsilon^{ac}   +  i \tilde{g} (X^a_{(j)}\,X^\dagger_{c\,(j)} \varepsilon^{cb} - \varepsilon^{bc} X^\dagger_{c\,(j-1)}\,X^a_{(j-1)}) - \frac{i \tilde{g}}{2} (X_{(j)}\,X^\dagger_{(j)} - X^\dagger_{(j-1)}X_{(j-1)} ) \varepsilon^{ab}\\
\delta \phi^a_{b\,(j)} = - \tilde{g} \varepsilon_{cb} \lambda^{ca}_{\,(j)} + \frac{\tilde{g}}{2} \delta^a_{b} \varepsilon_{cd}\,\lambda^{cd}_{\,(j)} \\
\delta X^a_{(j)} = - i \,\varepsilon^a_b\,\xi^b_{(j)}, \qquad \delta \xi^a_{(j)} =  \slashed{\mathcal{D}} X^b_{(j)} \varepsilon^a_b  + \frac13 X^b_{(j)} \slashed{\nabla} \varepsilon^a_b +  \phi^a_{b\,(j)} \varepsilon^b_c\, X^c_{(j)} - X^c_{(j)}\varepsilon^b_c\, \phi^a_{b\,(j+1)} \nonumber  \\ \label{hp-susy}
\delta X^\dagger_{a\,(j)} = -i \,\xi^\dagger_{b\,(j)} \varepsilon^b_a, \qquad  \delta \xi^\dagger_a = \slashed{\mathcal{D}} X^\dagger_{b\,(j)} \varepsilon^b_a+ \frac13 X^\dagger_{b\,{(j)}} \slashed{\nabla} \varepsilon^b_a  - \phi^b_{a\,(j+1)} \varepsilon^c_b \,X^\dagger_{c\,(j)} +  X^\dagger_{c\,(j)}  \varepsilon^c_b\, \phi^b_{a\,(j)}.
\end{eqnarray}

\section{\label{3} Classical supersymmetric monopole solutions}
In this section, we obtain the classical BPS bosonic monopole solutions to the theory constructed in \ref{22}. We first solve the vector multiplet fermion variation equations \eqref{sp-susy} for each node of the quiver and the hyper multiplet fermion variation equations \eqref{hp-susy} for each edge of the quiver; that is, we find non-trivial background fields $A_{m\,(j)}, \phi_{(j)}, X_{(j)}$ and an appropriate choice of supersymmetry parameters $\varepsilon_i$ that set the fermion variations to zero. The background fields of such a non-trivial solution to the fermion variation equations should then be checked if they solve the classical equations of motion derived from the action \eqref{SkinS} and \eqref{SintS}. We also aim to obtain a solution all along the RG flow, that is a solution independent of the the coupling $\tilde{g}$.

Our ansatz for the gauge field is
\begin{equation}\label{gaugefield}
A_{(j)} = \frac{\mathbf{H}}{2} (\pm 1 - \cos \theta)\,d\varphi,
\end{equation}
limiting ourselves to the special case where the magnetic charges for all nodes are identical.  We note that the quantity that appears often $\epsilon^{m n k}\, F_{m n \,(j)}$ is constant and non-zero only for $k = \tau$ with  $\epsilon^{m n \tau}\, F_{m n \,(j)} = \mathbf{H}$. 

First we study the vector multiplet fermion equations \eqref{sp-susy} which has two kinds of terms, viz. terms of order $\frac{1}{\tilde{g}}$ and terms at order $\tilde{g}$. The goal of finding solutions all along the flow means we have to set them separately to zero. 

\underline{\textbf{$\delta \lambda^{ab}_{(j)}$ at order $\frac{1}{\tilde{g}}$}:} We obtain from \eqref{sp-susy}
\begin{eqnarray}
\frac{i}{2} \epsilon^{m n k}\,F_{m n \,(j)} \gamma_k\,\varepsilon^{ab} - i \slashed{\mathcal{D}} \phi^b_{c\,(j)}\varepsilon^{ac} - \frac{2 i}{3} \phi^b_{c\,(j)} \slashed{\nabla} \varepsilon^{ac}+ \frac{i}{2} [{\phi^b_c}_{{(j)}}},{{\phi^c_d}_{{(j)}}] \varepsilon^{ad} = 0.
\end{eqnarray}
We work in a gauge (similar to \cite{Benna:2009xd}) where the adjoint fields $\phi_{(j)}$ are in the Cartan subalgebra and hence the commutators in the fourth term above vanish. We also assume that they are constant functions which means that the second term above vanishes. The third term simplifies after using the Killing spinor equation \eqref{ksp} and we obtain
\begin{eqnarray} \label{337}
\frac{i}{2} \mathbf{H} \,\gamma_\tau\,\varepsilon^{ab} +  i\,   \gamma_\tau\varepsilon^{ac}\,\phi^{~b}_{c\,(j)} = 0
\end{eqnarray}
$\delta \lambda^{ab}_{\,(j)}$ is in the reducible reprsentation $\mathbf{2}\times\mathbf{2} = \mathbf{1} + \mathbf{3}$ of $SU(2)_R$.  To isolate the $\mathbf{1}$ part i.e. the $SU(2)_R$ trace, we compute $\delta \lambda^{ab}_{\,(j)}\,\epsilon_{b a}$. The first term in \eqref{337}  being proportional to the supersymmetry parameter is $SU(2)_R$ traceless hence only the trace of the second term contributes to the $SU(2)_R$ trace of $\delta \lambda^{ab}_{\,(j)}$ at order $\frac{1}{\tilde{g}}$ and we get
\begin{equation}\label{-1trace}
\varepsilon_i\,\phi_{i\,(j)} = 0.
\end{equation}
We thus find that the non-trivial supersymmetry transformation parameter and the non-trivial background field $\phi_{(j)}$ are orthogonal.  Then we isolate the $\mathbf{3}$ part by  computing $\delta \lambda^{ab}_{\,(j)}\,\sigma_{i\,b a}$ and we get
\begin{equation}\label{-1triplet}
\frac{i}{2}\,\mathbf{H} \,\gamma_\tau  \varepsilon_i +  \epsilon_{i k l} \phi_{k\,(j)} \,\gamma_\tau \varepsilon_l = 0.
\end{equation}
We can solve \eqref{-1trace} by taking 
\begin{equation}
\varepsilon_3 = 0, \qquad \phi_{i~(j)} \sim \delta_{i3}
\end{equation}
which makes the equations \eqref{-1triplet}
\begin{equation}
\frac{\mathbf{H}}{2}\, \gamma_\tau\, \,i\varepsilon_1 - \phi_{3\,(j)}\,\gamma_\tau\, \varepsilon_2 = 0, \quad \frac{\mathbf{H}}{2}\, \gamma_\tau\, \,i\varepsilon_2 + \phi_{3\,(j)}\,\gamma_\tau\, \varepsilon_1 = 0
\end{equation}
easily solvable in two ways:
\begin{eqnarray}
(i)\qquad \phi_{3\,(j)} = - \frac{\mathbf{H}}{2}, \quad \varepsilon_1 - i \varepsilon_2 = 0 \\
(ii)\qquad \phi_{3\,(j)} =  \frac{\mathbf{H}}{2}, \quad \varepsilon_1 + i \varepsilon_2 = 0.
\end{eqnarray}
In the first case $(i)$, the preserved supersymmetry is $\varepsilon_1 + i \varepsilon_2$ and is the BPS solution and the second case $(ii)$, the preserved supersymmetry is $\varepsilon_1 - i \varepsilon_2$ and is the anti-BPS solution. In both cases, since one of the three supersymmetry parameters are preserved by the solution we have $\frac13$-BPS solutions.
To summarise the vector multiplet fermion equations at order $\frac{1}{\tilde{g}}$
are solved for $\eta = \pm1$ by 
\begin{eqnarray}\label{343}
A_{(j)} = \frac{\mathbf{H}}{2} (\pm 1 - \cos \theta)\,d\varphi, \quad \phi_{i\,(j)} = - \eta \frac{\mathbf{H}}{2}\,\delta_{i3}
\end{eqnarray}
 with $\varepsilon_1 + i \eta\,\varepsilon_2$ the preserved supersymmetry. Note that till this stage, we could have chosen $A_{(j)} \sim H_{(j)}$ and one could still have solved and obtained BPS and anti-BPS solutions with $\phi_{i\,(j)} = - \eta \frac{\mathbf{H_{(j)}}}{2}\,\delta_{i3}$.

\underline{\textbf{$\delta \lambda^{ab}_{(j)}$ at order $\tilde{g}$}:}  We obtain from \eqref{sp-susy}
\begin{eqnarray}\label{orderg}
\kappa_{(j)}   \,\phi^b_{c\,(j)}\varepsilon^{ac}   +   (X^a_{(j)}\,X^\dagger_{c\,(j)} \varepsilon^{cb} - \varepsilon^{bc} X^\dagger_{c\,(j-1)}\,X^a_{(j-1)}) - \frac{1}{2} (X_{(j)}\,X^\dagger_{(j)} - X^\dagger_{(j-1)}X_{(j-1)} ) \varepsilon^{ab} = 0
\end{eqnarray}
We analyse the above as before by considering the $\mathbf{1}$ and the $\mathbf{3}$ parts separately. The $SU(2)_R$ trace of the first term is  $\kappa_{(j)}\,\varepsilon_i\,\phi_{i(j)}$ which vanishes on using \eqref{-1trace} and the third term is traceless. The  second term gives
\begin{eqnarray}\label{+1trace}
(i) &\eta =1\qquad  (-W^\dagger_{(j)}Z^\dagger_{(j)} + Z^\dagger_{(j-1)} W^\dagger_{(j-1)})(\varepsilon_1 + i \varepsilon_2 ) = 0\ \nonumber \\
(ii) &\eta =-1\qquad (-Z_{(j)}W_{(j)} + W_{(j-1)} Z_{(j-1)})(\varepsilon_1 - i \varepsilon_2 ) = 0.
\end{eqnarray}
By isolating the $\mathbf{3}$ part of \eqref{orderg} one obtains the following equations
\begin{equation}\label{+1three}
2 \kappa_{(j)}\,\phi_{i\,(j)} = - X_{(j)}\sigma_i X^\dagger_{(j)} + X^\dagger_{(j-1)}\sigma_i X_{(j-1)}
\end{equation}
by which we see that the hyper multiplet scalars are already constrained by the levels and the magnetic charges in addition to other constraints that are yet to come from analysing the hyperino equations.

\underline{$\mathbf{\delta \xi^{a}_{(j)}}$:} We obtain from \eqref{hp-susy}
\begin{eqnarray}
\slashed{\mathcal{D}} X^b_{(j)} \varepsilon^a_b  + \frac13 X^b_{(j)} \slashed{\nabla} \varepsilon^a_b +  \phi^a_{b\,(j)} \varepsilon^b_c\, X^c_{(j)} - X^c_{(j)}\varepsilon^b_c\, \phi^a_{b\,(j+1)} = 0.
\end{eqnarray}
The last two terms cancel\footnote{There won't be a straight forward cancellation in the more general case  where the magnetic charges are different at each node, $A_{(j)} \sim H_{(j)}, ~ \phi_{i\,(j)} = - \eta \frac{\mathbf{H_{(j)}}}{2}\,\delta_{i3}$ \cite{nextwork}.} because in the special case we are working in, all the $\phi_{(j)}$'s are equal. After using the Killing spinor equation \eqref{ksp} the first two terms give a $\tau$ dependence $e^{\frac{\tau}{2}}$ to $X^2_{(j)}$  in the BPS case and to $X^1_{(j)}$ in the anti-BPS case, leaving the other hyper multiplet scalars unconstrained.   This functional dependence $X^2_{(j)} \sim e^{ \frac{\eta \tau}{2}}$ and $X^1_{(j)} \sim e^{-\frac{\eta \tau}{2}}$ is consistent with the equations of motion of the hyper multiplet scalars.

The only other equations of motion that remain to be satisfied by the so-far obtained background are for the gauge fields:
\begin{equation}\label{eomgauge}
\kappa_{(j)} \epsilon^{m n k}\,F_{m n\,(j)} = X_{(j)} \mathcal{D}^m X^\dagger_{(j)} - \mathcal{D}^m X_{(j)} X^\dagger_{(j)} - X^\dagger_{(j)} \mathcal{D}^m X_{(j)} +  X^\dagger_{(j-1)} \mathcal{D}^m X_{(j-1)}.
\end{equation}

To summarise so far, we have completely obtained the background gauge and adjoint scalar fields in \eqref{343} and we have several constraints on the hyper multiplet scalar fields: (i) equation \eqref{+1trace}, (ii) equation \eqref{+1three} (iii) the $\tau$ dependence that follow from the hyperino variations and (iv) the equation \eqref{eomgauge}. We will analyse these constraints and find solutions first for the three node quiver and with the experience gained thus, we can then generalize for a generic $n$-node quiver.

\subsection{\label{31}Three node quiver}
For $n = 3$, the constraint on the levels is 
\begin{equation}
\kappa_{(1)} + \kappa_{(2)} + \kappa_{(3)} = 0
\end{equation}
At least one of the levels needs to be negative. 

\underline{\textbf{one negative level}:} First we consider the case when one of the levels is negative say $\kappa_{(2)}$. For a positive semi-definite $\mathbf{H}=\mathop{\mathrm{diag}} \{q_1, q_2, \dots q_N \}$, we can solve the several constraints on the hyper multiplet scalars for BPS solutions in the following ways:
\begin{eqnarray}
\underline{\mathop{\mathrm{(i)}} \quad \eta=1}, \qquad Z_{(1)} = \sqrt{A}\,e^{-\frac{\tau}{2}}, \quad Z^\dagger_{(1)} = \sqrt{A}\,e^{\frac{\tau}{2}}, \quad W_{(1)} = 0, \quad W^\dagger_{(1)} = 0 \nonumber \\
Z_{(2)} = 0, \quad Z^\dagger_{(2)} = 0, \quad W_{(2)} = \sqrt{\mathbf{H}\, | \kappa_{(2)} | - A}\, e^{\frac{\tau}{2}}, \quad W^\dagger_{(2)} =  \sqrt{\mathbf{H}\, | \kappa_{(2)} | - A}\, e^{-\frac{\tau}{2}}, \nonumber \\
Z_{(3)} = \sqrt{A - \mathbf{H}\,  \kappa_{(1)}}\,e^{-\frac{\tau}{2}}, \quad Z^\dagger_{(3)} = \sqrt{A - \mathbf{H}\,  \kappa_{(1)}}\,e^{\frac{\tau}{2}}, \quad W_{(3)} = 0, \quad W^\dagger_{(3)} = 0 
\end{eqnarray}
where $A$ is a diagonal matrix with positive entries such that $q_\alpha \,\kappa_{(1)} \leq A_\alpha \leq q_\alpha |\kappa_{(2)} |$
\begin{eqnarray}
\underline{\mathop{\mathrm{(ii)}} \quad \eta=1}, \qquad Z_{(1)} = 0, \quad Z^\dagger_{(1)} = 0, \quad W_{(1)} = \sqrt{A}\,e^{\frac{\tau}{2}}, \quad W^\dagger_{(1)} = \sqrt{A}\,e^{-\frac{\tau}{2}} \nonumber \\
Z_{(2)} = 0, \quad Z^\dagger_{(2)} = 0, \quad W_{(2)} = \sqrt{A - \mathbf{H}\,  \kappa_{(2)}}\, e^{\frac{\tau}{2}}, \quad W^\dagger_{(2)} =  \sqrt{A - \mathbf{H}\,  \kappa_{(2)}}\, e^{-\frac{\tau}{2}}, \nonumber \\
Z_{(3)} = 0, \quad Z^\dagger_{(3)} = 0, \quad W_{(3)} = \sqrt{A + \mathbf{H}\,  \kappa_{(1)}} \, e^{\frac{\tau}{2}}, \quad W^\dagger_{(3)} = \sqrt{A + \mathbf{H}\,  \kappa_{(1)}} \,e^{-\frac{\tau}{2}}
\end{eqnarray}
where $A$ is a any diagonal matrix with positive entries.
\begin{eqnarray}
\underline{\mathop{\mathrm{(iii)}} \quad \eta=1}, \qquad Z_{(1)} = \sqrt{A}\,e^{-\frac{\tau}{2}}, \quad Z^\dagger_{(1)} = \sqrt{A}\,e^{\frac{\tau}{2}}, \quad W_{(1)} =0, \quad W^\dagger_{(1)} = 0 \nonumber \\
Z_{(2)} = \sqrt{A + \mathbf{H}\,  \kappa_{(2)}}\,e^{-\frac{\tau}{2}}, \quad Z^\dagger_{(2)} = \sqrt{A + \mathbf{H}\,  \kappa_{(2)}}\,e^{\frac{\tau}{2}}, \quad W_{(2)} = 0, \quad W^\dagger_{(2)} =  0, \nonumber \\
Z_{(3)} = \sqrt{A - \mathbf{H}\,  \kappa_{(1)}}\,e^{-\frac{\tau}{2}}, \quad Z^\dagger_{(3)} = \sqrt{A - \mathbf{H}\,  \kappa_{(1)}}\,e^{\frac{\tau}{2}}, \quad W_{(3)} = 0, \quad W^\dagger_{(3)} = 0
\end{eqnarray}
where $A$ is a  diagonal matrix with positive entries such that $A_\alpha \geq q_\alpha \kappa_{(1)}$ and $A_\alpha \geq q_\alpha |\kappa_{(2)}|$.

For a positive semi-definite $\mathbf{H}=\mathop{\mathrm{diag}} \{q_1, q_2, \dots q_N \}$, we can solve the several constraints on the hyper multiplet scalars for anti-BPS solutions in the following ways:
\begin{eqnarray}
\underline{\mathop{\mathrm{(i)}} \quad \eta=-1}, \qquad Z_{(1)} = 0, \quad Z^\dagger_{(1)} = 0, \quad W_{(1)} = \sqrt{A + \mathbf{H}\,  \kappa_{(1)}}\,e^{-\frac{\tau}{2}}, \quad W^\dagger_{(1)} = \sqrt{A + \mathbf{H}\,  \kappa_{(1)}}\,e^{\frac{\tau}{2}} \nonumber \\
Z_{(2)} = \sqrt{\mathbf{H}\,  \kappa_{(3)}  - A}\,e^{\frac{\tau}{2}}, \quad Z^\dagger_{(2)} = \sqrt{\mathbf{H}\,  \kappa_{(3)}  - A}\,e^{-\frac{\tau}{2}}, \quad W_{(2)} = 0, \quad W^\dagger_{(2)} =  0, \nonumber \\
Z_{(3)} = 0, \quad Z^\dagger_{(3)} = 0, \quad W_{(3)} = \sqrt{A}\,e^{-\frac{\tau}{2}}, \quad W^\dagger_{(3)} = \sqrt{A}\,e^{\frac{\tau}{2}}
\end{eqnarray}
where $A$ is a diagonal matrix with positive entries such that $A_\alpha \leq q_\alpha \kappa_{(3)}$.
\begin{eqnarray}
\underline{\mathop{\mathrm{(ii)}} \quad \eta=-1}, \qquad Z_{(1)} = 0, \quad Z^\dagger_{(1)} = 0, \quad W_{(1)} = \sqrt{A}\,e^{-\frac{\tau}{2}}, \quad W^\dagger_{(1)} = \sqrt{A}\,e^{\frac{\tau}{2}} \nonumber \\
Z_{(2)} = 0, \quad Z^\dagger_{(2)} = 0, \quad W_{(2)} = \sqrt{A + \mathbf{H}\,  \kappa_{(2)}}\, e^{-\frac{\tau}{2}}, \quad W^\dagger_{(2)} =  \sqrt{A + \mathbf{H}\,  \kappa_{(2)}}\, e^{\frac{\tau}{2}}, \nonumber \\
Z_{(3)} = 0, \quad Z^\dagger_{(3)} = 0, \quad W_{(3)} = \sqrt{A - \mathbf{H}\,  \kappa_{(1)}} \, e^{-\frac{\tau}{2}}, \quad W^\dagger_{(3)} = \sqrt{A - \mathbf{H}\,  \kappa_{(1)}} \,e^{\frac{\tau}{2}}
\end{eqnarray}
where $A$ is a diagonal matrix with positive entries such that $A_\alpha \geq q_\alpha\,\kappa_{(1)}$ and $A_\alpha \geq q_\alpha\,|\kappa_{(2)}|$
\begin{eqnarray}
\underline{\mathop{\mathrm{(iii)}} \quad \eta=-1}, \qquad Z_{(1)} = \sqrt{A}\,e^{\frac{\tau}{2}}, \quad Z^\dagger_{(1)} = \sqrt{A}\,e^{-\frac{\tau}{2}}, \quad W_{(1)} =0, \quad W^\dagger_{(1)} = 0 \nonumber \\
Z_{(2)} = \sqrt{A - \mathbf{H}\,  \kappa_{(2)}}\,e^{\frac{\tau}{2}}, \quad Z^\dagger_{(2)} = \sqrt{A - \mathbf{H}\,  \kappa_{(2)}}\,e^{-\frac{\tau}{2}}, \quad W_{(2)} = 0, \quad W^\dagger_{(2)} =  0, \nonumber \\
Z_{(3)} = \sqrt{A + \mathbf{H}\,  \kappa_{(1)}}\,e^{\frac{\tau}{2}}, \quad Z^\dagger_{(3)} = \sqrt{A + \mathbf{H}\,  \kappa_{(1)}}\,e^{-\frac{\tau}{2}}, \quad W_{(3)} = 0, \quad W^\dagger_{(3)} = 0
\end{eqnarray}
where $A$ is a any diagonal matrix with positive entries.

We have thus obtained when one of the levels is negative and for positive semi-definite $\mathbf{H}$, three families each of BPS and anti-BPS solutions, each family parametrised by a positive diagonal matrix  (with entries that are not unrestricted).

\underline{\textbf{two negative levels}:} Now we consider the case when two of the Chern-Simons levels are negative say $\kappa_{(1)}$ and $\kappa_{(2)}$. The BPS solutions are:
\begin{eqnarray}
\underline{\mathop{\mathrm{(i)}} \quad \eta=1}, \qquad Z_{(1)} = \sqrt{A}\,e^{-\frac{\tau}{2}}, \quad Z^\dagger_{(1)} = \sqrt{A}\,e^{\frac{\tau}{2}}, \quad W_{(1)} = 0, \quad W^\dagger_{(1)} = 0 \nonumber \\
Z_{(2)} = 0, \quad Z^\dagger_{(2)} = 0, \quad W_{(2)} = \sqrt{\mathbf{H}\, | \kappa_{(2)} | - A}\, e^{\frac{\tau}{2}}, \quad W^\dagger_{(2)} =  \sqrt{\mathbf{H}\, | \kappa_{(2)} | - A}\, e^{-\frac{\tau}{2}}, \nonumber \\
Z_{(3)} = \sqrt{A - \mathbf{H}\,  \kappa_{(1)}}\,e^{-\frac{\tau}{2}}, \quad Z^\dagger_{(3)} = \sqrt{A - \mathbf{H}\,  \kappa_{(1)}}\,e^{\frac{\tau}{2}}, \quad W_{(3)} = 0, \quad W^\dagger_{(3)} = 0 
\end{eqnarray}
where $A$ is a diagonal matrix with positive entries such that $ A_\alpha \leq q_\alpha |\kappa_{(2)} |$.
\begin{eqnarray}
\underline{\mathop{\mathrm{(ii)}} \quad \eta=1}, \qquad Z_{(1)} = 0, \quad Z^\dagger_{(1)} = 0, \quad W_{(1)} = \sqrt{A}\,e^{\frac{\tau}{2}}, \quad W^\dagger_{(1)} = \sqrt{A}\,e^{-\frac{\tau}{2}} \nonumber \\
Z_{(2)} = 0, \quad Z^\dagger_{(2)} = 0, \quad W_{(2)} = \sqrt{A - \mathbf{H}\,  \kappa_{(2)}}\, e^{\frac{\tau}{2}}, \quad W^\dagger_{(2)} =  \sqrt{A - \mathbf{H}\,  \kappa_{(2)}}\, e^{-\frac{\tau}{2}}, \nonumber \\
Z_{(3)} = 0, \quad Z^\dagger_{(3)} = 0, \quad W_{(3)} = \sqrt{A + \mathbf{H}\,  \kappa_{(1)}} \, e^{\frac{\tau}{2}}, \quad W^\dagger_{(3)} = \sqrt{A + \mathbf{H}\,  \kappa_{(1)}} \,e^{-\frac{\tau}{2}}
\end{eqnarray}
where $A$ is a any diagonal matrix with positive entries such that $A_\alpha \geq q_\alpha |\kappa_{(1)}|$.
\begin{eqnarray}
\underline{\mathop{\mathrm{(iii)}} \quad \eta=1}, \qquad Z_{(1)} = \sqrt{A}\,e^{-\frac{\tau}{2}}, \quad Z^\dagger_{(1)} = \sqrt{A}\,e^{\frac{\tau}{2}}, \quad W_{(1)} =0, \quad W^\dagger_{(1)} = 0 \nonumber \\
Z_{(2)} = \sqrt{A + \mathbf{H}\,  \kappa_{(2)}}\,e^{-\frac{\tau}{2}}, \quad Z^\dagger_{(2)} = \sqrt{A + \mathbf{H}\,  \kappa_{(2)}}\,e^{\frac{\tau}{2}}, \quad W_{(2)} = 0, \quad W^\dagger_{(2)} =  0, \nonumber \\
Z_{(3)} = \sqrt{A - \mathbf{H}\,  \kappa_{(1)}}\,e^{-\frac{\tau}{2}}, \quad Z^\dagger_{(3)} = \sqrt{A - \mathbf{H}\,  \kappa_{(1)}}\,e^{\frac{\tau}{2}}, \quad W_{(3)} = 0, \quad W^\dagger_{(3)} = 0
\end{eqnarray}
where $A$ is a  diagonal matrix with positive entries such that $A_\alpha \geq q_\alpha |\kappa_{(2)}|$.

The anti-BPS solutions are:
\begin{eqnarray}
\underline{\mathop{\mathrm{(i)}} \quad \eta=-1}, \qquad Z_{(1)} = 0, \quad Z^\dagger_{(1)} = 0, \quad W_{(1)} = \sqrt{A + \mathbf{H}\,  \kappa_{(1)}}\,e^{-\frac{\tau}{2}}, \quad W^\dagger_{(1)} = \sqrt{A + \mathbf{H}\,  \kappa_{(1)}}\,e^{\frac{\tau}{2}} \nonumber \\
Z_{(2)} = \sqrt{\mathbf{H}\,  \kappa_{(3)}  - A}\,e^{\frac{\tau}{2}}, \quad Z^\dagger_{(2)} = \sqrt{\mathbf{H}\,  \kappa_{(3)}  - A}\,e^{-\frac{\tau}{2}}, \quad W_{(2)} = 0, \quad W^\dagger_{(2)} =  0, \nonumber \\
Z_{(3)} = 0, \quad Z^\dagger_{(3)} = 0, \quad W_{(3)} = \sqrt{A}\,e^{-\frac{\tau}{2}}, \quad W^\dagger_{(3)} = \sqrt{A}\,e^{\frac{\tau}{2}}
\end{eqnarray}
where $A$ is a diagonal matrix with positive entries such that $q_\alpha |\kappa_{(1)}| \leq A_\alpha \leq q_\alpha \kappa_{(3)}$.
\begin{eqnarray}
\underline{\mathop{\mathrm{(ii)}} \quad \eta=-1}, \qquad Z_{(1)} = 0, \quad Z^\dagger_{(1)} = 0, \quad W_{(1)} = \sqrt{A}\,e^{-\frac{\tau}{2}}, \quad W^\dagger_{(1)} = \sqrt{A}\,e^{\frac{\tau}{2}} \nonumber \\
Z_{(2)} = 0, \quad Z^\dagger_{(2)} = 0, \quad W_{(2)} = \sqrt{A + \mathbf{H}\,  \kappa_{(2)}}\, e^{-\frac{\tau}{2}}, \quad W^\dagger_{(2)} =  \sqrt{A + \mathbf{H}\,  \kappa_{(2)}}\, e^{\frac{\tau}{2}}, \nonumber \\
Z_{(3)} = 0, \quad Z^\dagger_{(3)} = 0, \quad W_{(3)} = \sqrt{A - \mathbf{H}\,  \kappa_{(1)}} \, e^{-\frac{\tau}{2}}, \quad W^\dagger_{(3)} = \sqrt{A - \mathbf{H}\,  \kappa_{(1)}} \,e^{\frac{\tau}{2}}
\end{eqnarray}
where $A$ is a diagonal matrix with positive entries such that $q_\alpha\,|\kappa_{(2)}| \leq A_\alpha$. 
\begin{eqnarray}
\underline{\mathop{\mathrm{(iii)}} \quad \eta=-1}, \qquad Z_{(1)} = \sqrt{A}\,e^{\frac{\tau}{2}}, \quad Z^\dagger_{(1)} = \sqrt{A}\,e^{-\frac{\tau}{2}}, \quad W_{(1)} =0, \quad W^\dagger_{(1)} = 0 \nonumber \\
Z_{(2)} = \sqrt{A - \mathbf{H}\,  \kappa_{(2)}}\,e^{\frac{\tau}{2}}, \quad Z^\dagger_{(2)} = \sqrt{A - \mathbf{H}\,  \kappa_{(2)}}\,e^{-\frac{\tau}{2}}, \quad W_{(2)} = 0, \quad W^\dagger_{(2)} =  0, \nonumber \\
Z_{(3)} = \sqrt{A + \mathbf{H}\,  \kappa_{(1)}}\,e^{\frac{\tau}{2}}, \quad Z^\dagger_{(3)} = \sqrt{A + \mathbf{H}\,  \kappa_{(1)}}\,e^{-\frac{\tau}{2}}, \quad W_{(3)} = 0, \quad W^\dagger_{(3)} = 0
\end{eqnarray}
where $A$ is a a diagonal matrix with positive entries such that $q_\alpha\,|\kappa_{(1)}| \leq A_\alpha$.

We thus have again obtained three families of solutions for each of the BPS and anti-BPS cases. For the full solution one should supplement the hyper multiplet scalar backgrounds of this section \ref{31} with the backgrounds for the gauge fields and the adjoint scalars \eqref{343}.

\subsection{\label{32}The $n$-node quiver}
We are looking  for solutions for diagonal $Z$ and $W$. From the $\phi_1$ and $\phi_2$ equations of \eqref{+1three} we get:
\begin{equation} \label{b1}
Z_{(i)\,\alpha}W_{(i)\,\alpha} = d_{\alpha}.
\end{equation}
From the $\phi_3$ equation of \eqref{+1three} we get :
\begin{equation} \label{b2}
|Z_{(i)\,\alpha}|^2 - | W_{(i)\alpha}|^2 = |Z_{(1) \,\alpha}|^2 - | W_{(1)\alpha}|^2 +\eta \,q_{\alpha}\,\Sigma^i_{j=2} \kappa_{(j)} \; \; \; (\forall i\geq 2) 
\end{equation}
Suppressing the $\alpha$ index and further using the notation $|Z_{(i)\alpha}|^2 \equiv u_i$, $|W_{(i)\alpha}|^2 \equiv v_i$ and $\Sigma^i_{j=2} \kappa_{(j)} \equiv K_i$, we can write the above equation in a compact form:
\begin{equation}\label{b3}
u_i -v_i = u_1 -v_1 + \eta \,q \,K_i
\end{equation} 

{\bf Case I: $d_{\alpha}\neq 0$}
\\
In this case, using equations \eqref{b1} and \eqref{b2}, we get :
\begin{equation}
(u_{i}- u_1)\,(u_1 u_i +|d|^2) =\eta \,K_i\, q \,u_1\,u_i
\end{equation}
The solution for this is:
\begin{equation}\label{b5}
u_i = \frac{u^2_1 -|d|^2 +\eta \,K_i\,u_1 \pm \sqrt{(|d|^2-u^2_1-\eta \,K_i \,q \,u_1)^2 +4|d|^2\,u^2_1}}{2u_1}
\end{equation}
These seem to exist for any value of $\kappa_{(i)}$. These are a moduli space of solutions classified by $Z_{(1)\alpha}$, $\arg(Z_{{(i)}\alpha})$ and $d_\alpha$. 

{\bf Case II: $d_{\alpha}= 0$}

Lets look at the three node case :

We can choose either $u_1 =0$ or $v_1 =0$. Lets look at the case where $u_1=0$ first. 
We can have the following possibilites, [$u_2=u_3=0$], [$v_2=v_3=0$], [$u_2=v_3=0$], [$v_2=u_3=0$]. 
\\

{\bf case 1: [$u_2=u_3=0$]}
\begin{equation}
v_2 =-u_1 -\eta \,q\,K_2, \; \; v_3 = - u_1 -\eta\, q\,K_3
\end{equation}

For $\eta=+1$, solution exists iff $K_2<0$ and $K_3<0$ and $u_1 <  q . \min(|K_2|,|K_3|)$.

For $\eta =-1$, solutions exist iff $K_1>0$ and $K_3>0$ and $u_1< q. \min(K_2, K_3)$.
\\

{\bf case 2: [$v_2=v_3=0$]}
\begin{equation}
u_2 =u_1 +\eta \, q\,K_2, \; \; u_3 = u_1 +\eta\, q\,K_3
\end{equation}

For $\eta=+1$, solution exists if $K_2, K_3>0$. They also exist if both of them are negative provided $u_i> q.\max(|K_2|,|K|_3)$. For one of them(say $K_2$)  negative, solutions exist if $u_1>q.|K_2|$.

For $\eta =-1$, solutions always exist for  $K_2<0, K_3<0$. For  $K_2>0, K_3>0$ solutions exist provided $u_1< q. \max(K_2, K_3)$. For one of the two (say $K_2$) positive, solutions again exist provided $u_1>q.|K_2|$. 
\\

{\bf case 3: [$u_2=v_3=0$]}
\begin{equation}
v_2 =-u_1 -\eta \,q\,K_2, \; \; u_3 = u_1 + \eta\, q\,K_3
\end{equation}

For $\eta=+1$, solution exists iff $K_2<0$. Solutions always exist if $K_3>0$. For $K_3<0$, solutions exist provided $u_1>q.|K_3|$.

 For $\eta=-1$, solution exists iff $K_2>0$. Solutions always exist if $K_3<0$. For $K_3>0$, solutions exist provided $u_1>q.K_3$.
\\

{\bf case 4: [$v_2=u_3=0$]}
\begin{equation}
u_2 =u_1 +\eta \,q\,K_2, \; \; v_3 = -u_1 -\eta\, q\,K_3
\end{equation}

For $\eta=+1$, solution exists iff $K_3<0$. Solutions always exist if $K_2>0$. For $K_2<0$, solutions exist provided $u_1>q.|K_2|$.

 For $\eta=-1$, solution exists iff $K_3>0$. Solutions always exist if $K_2<0$. For $K_2>0$, solutions exist provided $u_1>q.K_2$.
\subsubsection{Solutions for general $n$}
Since equations \eqref{b1} and \eqref{b2} hold separately for each $\alpha$, we can choose $d_{\alpha}=0$ or $d_{\alpha}\neq 0$, separately for each $\alpha$, and solve them separately. The discussions will hold separately for each case. 

We can also find general solutions for those $\alpha$ for which $d_{\alpha}=0$ for the general $n$-node case. We can then get a complete classification of solutions in the general case. This is done as follows.

From equation \eqref{b3}, it turns out that instead of dealing with $\kappa_{(i)}$ the classification is much simpler using $K_i$.  Lets take $v_1 =0$. Without loss of generality, lets choose a certain set of $u_i =0$ and the complementary set of $v_i=0$. Lets first concentrate on the set $v_i=0$. For all such cases:
\begin{equation}
u_i = u_1 +\eta \,q\,K_i
\end{equation}
Positivity of $u_i$ implies that:
\begin{equation}
u_1 +\eta \,q\,K_i>0
\end{equation} 
This means that for $\eta =+1$ and all $K_i>0$ there is always a solution for any value of $u_1$. If however, there is a subset of $K_i<0$, which we denote by the set $[K_{-}]$, then $u_1 > q.\max[K_{-}]$. Now lets look at the complementary set with $u_i=0$. From equation \eqref{b3} it follows that 
\begin{equation}
v_i = -u_1 -\eta\, q\,K_i
\end{equation}
Again for $\eta=+1$, we get the condition:
\begin{equation}
u_1 +\eta\, q\,K_i <0
\end{equation}
Its clear that there are no solutions if any of these $K_1>0$. Thus solutions exist iff $K_1<0$. Moreover in this case, $u_1<q.\min[|K_i|]$

To summarise, lets assume that we have a general $n$ node system, with some of the $K_i>0$, which we denote by the set $[K_{+}]$ and the remaining $K_i<0$, which we denote by the set $[K_{-}]$. We are looking for  solutions corresponding to choosing some of the $u_i=0$ and the remaining $v_i=0$. 

The most general solution corresponds to first choosing a subset of the $[K_{-}]$ set, and assigning  $u_i=0$ there. Lets label this subset as $[K_{u-}]$. We denote the complement of this subset within $[K_{-}]$ as $ [K_{v-}]$. We then need to put the further condition that within this set, $u_1< q.\min[|K_{u-}|]$. 

On the remaining nodes, $K_i$ either belongs to the set $[K_{+}]$ or the set $[K_{v-}]$. On these nodes the $v_i=0$. The $u_1$ satisfies the constraint that $u_1 > q.\max[K_{v-}]$. Thus we get 
\begin{equation}\label{b14}
q.\min[|K_{u-}|]>u_1 > q.\max[K_{v-}]
\end{equation}

Thus the moduli space of the $d=0$ solutions, in a given theory, with a specific set of $[K_{+}]$ and $[K_{-}]$, is parametrized by the number of nodes on which $u_i=0$ and solutions of $u_1$ satisfying equation \eqref{b14}. 

A similar analysis can be done for the $\eta=-1$ case and we obtain
\begin{equation}\label{b15}
q.\min[|K_{u+}|]>u_1 > q.\max[K_{v+}]
\end{equation}

This is the most general solution in the $d=0$ branch. Taken along with the solution in equation \eqref{b5} for the $d\neq 0$ branch, we have ${\emph all}$ the solutions in the general $n$ node case.

\section{\label{4} Conclusion and Outlook}

In this paper first we have constructed the $\mathcal{N}=3$ UV completion of the $\mathcal{N} = 3$
Jafferis-Tomasiello super conformal quiver gauge theory. $\mathcal{N}=3$ supersymmetry and $SU(2)_R$ symmetry are preserved all along the flow. Starting from the superspace action, we have obtained the component action written in terms of $SU(2)_R$ symmetry multiplets. We have also obtained explicitly the supersymmetry transformations that leave the action invariant. We have done all the above both in $\mathbf{R}^{1,2}$ and in $\mathbf{R}\times \mathbf{S}^2$.

We have obtained classical monopole solutions which exist all along the RG flow. The classical solutions are interesting in their own right: they come in families, as part of spaces of solutions. It would be interesting to develop these methods to obtain more generic solutions such as with the $A_{(j)} = \frac{\mathbf{H}_{(j)}}{2} (\pm 1 - \cos \theta)\,d\phi, $.

This paper is meant to be the first part to the follow-up work \cite{nextwork} where we will study the monopole operators in the IR SCFT via the monopole solutions of the present paper. This is the BKK setup \cite{Benna:2009xd} and we hope to use this to study monopole operators in more generic $\mathcal{N} = 3$ SCFTs such as the ones in \cite{Crichigno:2017rqg}.

\begin{center}
\textbf{Acknowledgments}
\end{center}
We thank Ashoke Sen for comments and encouragement through the course of this work. Bobby Ezhuthachan would like to thank the NISER theory group for warm hospitality during visits to NISER when some part of this work was done.

\appendix
\section{\label{A} Notation and conventions}
We use identical conventions in every respect (the spinor conventions, the gamma matrices, the summation convention NE-SW, the left-side raising and lowering convenctions, the $SU(2)_R$ conventions etc) to \cite{Benna:2009xd}. The component expansion of the $\mathcal{N}=2$ superfields in the $\mathcal{N}=3$ vector multiplet are
\begin{multline}
\mathcal{V}_{(j)} = 2 i \theta \overline{\theta} \sigma_{(j)} - 2 \theta \gamma^m \overline{\theta} A_{m\,(j)}+ \sqrt{2} i \theta^2 \overline{\theta} \,{\chi^\dagger_\sigma}_{{(j)}} - \sqrt{2} i \overline{\theta}^2 \theta \,{\chi_\sigma}_{{(j)}} + \theta^2\,\overline{\theta}^2\,D_{(j)},
\\
{\Phi}_{{(j)}}={\phi}_{{(j)}}(x_L) +\sqrt{2}{\theta}{\chi_\phi}_{{(j)}}(x_L) +{\theta}^2 {F_\phi}_{{(j)}}(x_L), \nonumber \\
\overline{\Phi}_{{(j)}}={\phi^\dagger}_{{(j)}}(x_R) -\sqrt{2}\overline{\theta}{\chi^\dagger_\phi}_{{(j)}}(x_R) -\overline{\theta}^2 {F^\dagger_\phi}_{{(j)}}(x_R)
\end{multline}
and for the $\mathcal{N}=2$ superfields in the $\mathcal{N}=3$ hyper multiplet are
\begin{multline}
{\mathcal{Z}}_{{(j)}}={Z}_{{(j)}}(x_L) +\sqrt{2}{\theta}{\zeta}_{{(j)}}(x_L) +{\theta}^2 {F}_{{(j)}}(x_L), \nonumber \\
\overline{\mathcal{Z}}_{{(j)}}={Z^\dagger}_{{(j)}}(x_R) -\sqrt{2}\overline{\theta}{\zeta^\dagger}_{{(j)}}(x_R) -\overline{\theta}^2 {F^\dagger}_{{(j)}}(x_R) \\
{\mathcal{W}}_{{(j)}}={W}_{{(j)}}(x_L) +\sqrt{2}{\theta}{\omega}_{{(j)}}(x_L) +{\theta}^2 {G}_{{(j)}}(x_L) \nonumber \\
\overline{\mathcal{W}}_{{(j)}}={W^\dagger}_{{(j)}}(x_R) -\sqrt{2}\overline{\theta}{\omega^\dagger}_{{(j)}}(x_R) -\overline{\theta}^2 {G^\dagger}_{{(j)}}(x_R)
\end{multline}
where $x^m_L = x^m - i \, \theta \gamma^m \overline{\theta}$ and $x^m_R = x^m + i \, \theta \gamma^m \overline{\theta}$.

\end{document}